\documentclass[12pt,a4paper]{article}
\newcommand{\p}{\partial}
\newcommand{\calF}{\mathcal{F}}
\newcommand{\calO}{\mathcal{O}}
\newcommand{\calH}{\mathcal{H}}
\newcommand{\calL}{\mathcal{L}}
\newcommand{\VT}{e^{-\frac{1}{4} T^2}}
\newcommand{\vt}{e^{-T^2/4}}
\newcommand{\intdx}{\int d^{p+1}x}
\newcommand{\nn}{\nonumber\\}
\newcommand{\ep}{\epsilon}
\newcommand{\al}{\alpha}
\newcommand{\sgn}{\mathrm{sgn}}
\newcommand{\erf}{\mathrm{erf}}
\newcommand{\erfc}{\mathrm{erfc}}
\newcommand{\const}{\mathrm{const}}
\title{Rolling down to D-brane and tachyon matter}
\author{
{\sc Akira Ishida}\footnote{e-mail:ishida@eken.phys.nagoya-u.ac.jp}~
and
{\sc Shozo Uehara}\footnote{e-mail:uehara@eken.phys.nagoya-u.ac.jp}
\vspace{3mm}\\
{\sl Department of Physics, Nagoya University,}\\
{\sl Chikusa-ku, Nagoya 464-8602, Japan}}
\date{}

\begin{document}
\maketitle
\vspace{-80mm}
\begin{flushright}
DPNU-03-01\\ hep-th/0301179\\ January 2003
\end{flushright}
\vspace{55mm}

\maketitle
\begin{abstract}
We investigate the spatially inhomogeneous decay of an unstable
D-brane and construct an asymptotic solution which describes a
codimension one D-brane and the tachyon matter in boundary string
field theory.
In this solution, the tachyon matter exists around the
lower-dimensional D-brane.
\end{abstract}
\section{Introduction}
Time dependent solutions of tachyon have been vigorously investigated
recently. One of the purpose is to study the decay of an unstable
D-brane. It was shown in \cite{Sen1,Sen2} that unstable D-branes
become the pressure-less gas with non-zero energy density at late
time, which is called tachyon matter, if the tachyon rolls down to
the bottom of the potential homogeneously.
While the time dependent solution is described by the boundary
state, effective field theory approaches give similar results.
The Born-Infeld type effective field theory \cite{Ga}-\cite{Kl}
reproduces the known results of the rolling tachyon \cite{Sen3,Sen5}.
In \cite{ST,Mi}, the spatially homogeneous decay was analyzed by
using boundary string field theory (BSFT) \cite{Wi1}--\cite{NP}.
They gave the asymptotic solution $\dot{T}\sim \pm 1$,
which reproduces the property of the tachyon matter.
Since the unstable D$p$-brane can decay into D$(p-1)$-brane, it is
interesting to consider the spatially inhomogeneous tachyon which
represents the time evolution of the kink solution. In this paper, we
study the behavior of the spatially inhomogeneous decay of an unstable
D-brane in BSFT and give an asymptotic solution.
As we will see later, this solution asymptotically approaches
$\dot{T}\sim \sgn(x^0)\quad$($\sgn(\cdot)$: sign-function),
and the lower-dimensional D-brane appears with the tachyon matter.

This paper is organized as follows. In section \ref{sec:ID}, we
consider the inhomogeneous decay of an unstable D-brane and construct
the asymptotic solution which describes both the lower-dimensional
D-brane and the tachyon matter. Then we briefly discuss how the
energy concentrates to make the lower-dimensional D-brane in section
\ref{sec:EF}.
The final section is devoted to conclusion and discussion.

\section{Inhomogeneous decay of the unstable D-brane}\label{sec:ID}
In this section, we consider the late time behavior of the decaying
D-brane into the lower-dimensional D-brane in
BSFT.\footnote{Inhomogeneous rolling tachyon has been discussed also
in \cite{Sen4}-\cite{CF}.}
The action is given by\footnote{We set $\al'=2$ for simplicity and
the signature of the metric is taken to be $\{-,+,+,\cdots,+\}$.
This action is exact as long as $\p_\mu\p_\nu T=0$ and higher
derivative corrections are difficult to examine in BSFT.
Since some reliable results have been obtained in \cite{ST,Mi} and
other papers, we neglect the higher derivative terms and proceed to
investigate.}\cite{KMM2}
\begin{equation}
 S=-T_p \intdx\, \VT\, \calF
	(\eta^{\mu\nu} \p_\mu T \p_\nu T)\,,\label{action}
\end{equation}
where
\begin{equation}
 \calF(x)=\frac{x\,4^x\, \Gamma(x)^2}{2\Gamma(2x)}\,,
\end{equation}
and $T_p$ is the tension of a non-BPS D-brane.
The energy-momentum tensor obtained from this action is
\begin{equation}
 T_{\mu\nu}= T_p\,\VT\left[\,2\,\p_\mu T \p_\nu T\, \calF'
        - \eta_{\mu\nu}\,\calF\,\right]. \label{EM}
\end{equation}
Note that the energy-momentum conservation and the equation of motion
are related as
\begin{eqnarray}
 \p_\mu T^{\mu\nu} &=& T_p\,\p^\nu T \left\{
   2\p_\mu (\VT \p^\mu T \calF')+\frac{T}{2}\,\VT\calF \right\}\nn
   &=& -\p^\nu T\left\{\p_\mu\left(\frac{\p\calL}{\p(\p_\mu T)}\right)
    -\frac{\p\calL}{\p T} \right\}.
\end{eqnarray}

In order that the codimension one D-brane,\footnote{We take
the codimension one D-brane to be localized transverse to
$x^1$-direction.} i.e. kink solution, will be developed finally, we
set the initial condition at $x^0=0$ as
\begin{equation}
 T>0 \hspace{3ex} (x^1>0), \hspace{4ex}
 T(x^1)=-T(-x^1), \hspace{4ex}
 \dot{T}=0 \,, \label{IC}
\end{equation}
and we also impose that the initial value of the tachyon field is very
small, i.e. $|T| \ll 1$ at $x^0=0$. Having imposed the initial
conditions, we expect that $T\to\infty$ for $x^1>0$ and
$T\to -\infty$ for $x^1<0$ at late time and the kink solution is
dynamically generated. Due to the energy conservation, it is natural
to expect that some non-zero energy density remains in the infinite
region as well.\footnote{A possibility that the energy
dissipates to infinity is excluded as will be shown in section
\ref{sec:EF}.}

Since we are interested in the inhomogeneity transverse to the
codimension one D-brane,
we put $\p_i T=0$ for $i=2,\cdots,p$. Then the action becomes
\begin{equation}
 S=-T_p \intdx\, \VT\, \calF (z)\,,
\end{equation}
where
\begin{equation}
  z = -\dot{T}^2+T'^2
	\qquad(\dot{T}\equiv\p_0T,~T'\equiv\p_1 T)\,.
\end{equation}
The hamiltonian density is
\begin{equation}
 \calH =T^{00}=T_p\, \VT \left(\calF(z) +2 \dot{T}^2
 \calF'(z)\right)\,.\label{T00}
\end{equation}
Let us investigate the equation of motion,
\begin{equation}
 \VT \left( \p^\mu T \p_\mu z \,\calF''(z) +\p^\mu \p_\mu T
 \,\calF'(z) -\frac{1}{2}\,T z\,\calF'(z) +\frac{1}{4}\,T\calF(z)
 \right)=0. \label{EoM0}
\end{equation}
We assume that $z$ does not cross the singularity at $z=-1$
and restrict ourselves to $z>-1$, where $\calF$ is analytic.

First, let us consider that $z$ goes to some finite value other than
$-1$, i.e. does not hit a singularity as $x^0 \to\infty$.
Since we treat the case where $|T|\to\infty$ as $x^0\to\infty$,
the leading contribution to the equation of motion (\ref{EoM0}) leads
to (cf. ref.\cite{HH})
\begin{equation}
 D(z)=\calF(z)-2z\calF'(z)=0. \label{EoM1}
\end{equation}
This equation has no solutions for finite values of $z$ because
$D(z)$ is positive in $-1<z\leq0$ \cite{ST} and it is also
positive for $z>0$ since it can be written in that region as
\begin{equation}
  D(z)= 2\calF(z)\,\int^1_0 \frac{t^{2z}}{(1+t)^2} dt\qquad(z>0)\,.
\end{equation}
In the case of $z\to\infty$, we obtain the same equation
as eq.(\ref{EoM1})
because the asymptotic forms of $\calF(y)$ and $\calF'(y)$ for the
argument $y$ being large are given by
\begin{equation}
 \calF(y)=\sqrt{\pi y}+\calO(y^{-1/2})\,, \hspace{3ex}
 \calF'(y)=\frac{1}{2}\,\sqrt{\frac{\pi}{y}}+\calO(y^{-3/2})\,.
 \label{asympN0}
\end{equation}
And we see that $z\to\infty$ is a solution for eq.(\ref{EoM1}).

Next, we proceed to the case that $z\to-1$ as $x^0\to\infty$.
We may write $|\dot{T}|=1-\calO(\delta)$, $|\delta|\ll1$ at late
time, and assume $T'=\calO(\delta)$.
Then we see that $z+1=\calO(\delta)$ and $\dot{z}=\calO(\delta)$.
Since $\calF(y)$, $\calF'(y)$ and $\calF''(y)$ behave near $y=-1$ as
\begin{equation}
 \calF(y) \sim \frac{-1}{2(y+1)}\,, \hspace{3ex}
 \calF'(y) \sim \frac{1}{2(y+1)^2}\,, \hspace{3ex}
 \calF''(y) \sim \frac{-1}{(y+1)^3}\,,\label{Fasym}
\end{equation}
the leading contribution of eq.(\ref{EoM0}) in this case becomes
\begin{equation}
 2\dot{T} \dot{z} \,\calF''(z) -Tz\,\calF'(z)=0,
\end{equation}
which leads to, due to eq.(\ref{Fasym}),
\begin{equation}
 \frac{\dot{z}}{z+1}=\frac{Tz}{4\dot{T}}
	=\frac{T(-\dot{T}^2+{T'}^2)}{4\dot{T}}
 \simeq -\frac{T\dot{T}}{4}\,.
\end{equation}
Solving this equation, we obtain the leading term of $z+1$ as
\begin{equation}
 z+1=g(x^1)\, e^{-\frac{1}{8}T^2},\label{z1asym}
\end{equation}
where $g(x^1)$ is a function of $x^1$.\footnote{
The $g(x^1)=\const$ case corresponds to the $\ep(x^0)$ in \cite{ST}.}
Note that eqs.(\ref{z1asym}) and (\ref{T00}) implies finite energy
density at late time.
We stress here that the equation of motion requires that $z$ goes to
either $-1$ or infinity as $x^0\to\infty$, which generally holds
when we consider the rolling tachyon.
The static solution which represents the codimension one D-brane is
$T=ux^1$ ($u\to\infty$) \cite{KMM2} and the spatially homogeneous
solution for the rolling tachyon is $|\dot{T}| \to 1$ as
$x^0\to\infty$ \cite{Sen2}. Thus, considering the above analysis of
the equation of motion, we require that the asymptotic solution
satisfies
\begin{eqnarray}
 &&T=0, \hspace{6ex} T' \to \infty \hspace{7ex}\,(x^1=0)\,, \nn
 &&\dot{T} \to \sgn(x^1), \hspace{3ex} T' \to 0
 \hspace{5ex} (x^1 \neq 0)\,. \label{Tasym}
\end{eqnarray}

Let us construct the solution describing the decay of
an unstable D-brane into the lower-dimensional one.
We expect the asymptotic solution for $T$ as
\begin{equation}
 T=\sigma(x) \left(x^0 -\ep(x)\right)\qquad
	 \left(x = (x^0,x^1)\right)\,,\label{TP}
\end{equation}
where $\ep(x)$ is a small perturbation and $\sigma(x)$
is a function which satisfies
\begin{equation}
 \sigma(x) \sim \sgn(x^1) \hspace{3ex} (x^0 \to \infty)\,.
	\label{asymsigma}
\end{equation}

First, we consider the asymptotic behavior of the tachyon
for $x^1 \ne 0$.
In this region, $|T| \sim x^0$ and hence $z\to-1$.
We have
\begin{eqnarray}
 z+1 &=& 1-\left\{\sigma\,(1-\dot{\ep})
    +\dot{\sigma}\,(x^0-\ep)\right\}^2
    +\left\{\sigma'\,(x^0-\ep) -\sigma\,\ep'\right\}^2 \nn
 &\simeq& 1-\sigma^2 +2 \dot{\ep} -2(\sgn(x^1)x^0)\dot{\sigma}\nn
 &\simeq& 2(1-|\sigma|) +2 \dot{\ep} -2(\sgn(x^1)x^0)\dot{\sigma}\,,
\end{eqnarray}
where use has been made of eq.(\ref{asymsigma}) and we have assumed
that $\dot{\sigma}$ and $\sigma'$ are small.
We also require, at late time,
\begin{equation}
 |\dot{\ep}|\gg |(1-|\sigma|)|\qquad\mbox{and}\qquad
	|\dot{\ep}|\gg x^0\,|\dot{\sigma}|\,.
\label{epsilon}
\end{equation}
Thus, in order to satisfy eq.(\ref{z1asym}), or obtain the finite
energy solution, $\ep(x)$ is given by
\begin{equation}
 \ep(x)=f(x^1) \int_\infty^{x^0}dy\,e^{-\frac{1}{8}y^2}
  = -\sqrt{2\pi} f(x^1)\, \erfc (x^0/\sqrt{8})\,,\label{eq:ep}
\end{equation}
where $f(x^1)$ is a function of $x^1$ and we assume that it does not
vary so much. Then we have
\begin{equation}
 \VT \calF(z) \to 0, \hspace{3ex}
 \VT \calF'(z) \to \frac{1}{8f^2}\,. \label{region1}
\end{equation}
We shall proceed to $\sigma(x)$.
{}From eqs.(\ref{z1asym}), (\ref{asymsigma}) and (\ref{epsilon}),
the difference, $\sigma(x)-\sgn(x^1)$, should damp more rapidly
than $\exp(-(x^0)^2/8)$ at late time. Thus we find that $\sigma(x)$
can be given by
\begin{equation}
 \sigma(x)=\frac{2}{\sqrt{\pi}} \int_0^{c(x^0)^{1+\alpha} x^1}
    dy\,e^{-y^2} = \erf (c(x^0)^{1+\alpha} x^1)\,,
   \quad(\alpha>0) \label{signf}
\end{equation}
where $\alpha$ is a positive constant and $c$ is a constant, which can
be taken positive without lost of generality. Although we write
the asymptotic form of $T$ as eq.(\ref{TP}), note that $T$ is
correct up to $o(e^{-(x^0)^2/8})$ except the infinitesimally
small region around $x^1=0$.

Next, we consider the behavior of the tachyon at the remaining point,
i.e. $x^1=0$ as $x^0 \to\infty$ with the above $\epsilon(x)$
(\ref{eq:ep})  and $\sigma(x)$ (\ref{signf}).
At this point we first notice $T=0$ due to eq.(\ref{signf}).
Also we find that $T'\simeq 2c(x^0)^{2+\alpha}/\sqrt{\pi}\to\infty$
while $\dot{T}=0$, and hence $z\to\infty$.
The asymptotic forms of $\calF(y)$ and $\calF'(y)$ for
the argument $y$ being large are given in eq.(\ref{asympN0}).
Hence at exactly $x^1=0$, we find the following behavior for
$x^0\to\infty$,
\begin{equation}
 \VT \calF(z) \to \infty, \hspace{3ex}
 \VT \calF'(z) \to 0\,. \label{region3}
\end{equation}
Let us calculate more carefully. In the infinitesimally small region
around $x^1=0$ for $x^0\gg1$, $T$ in eq.(\ref{TP}) and its
derivatives are evaluated as
\begin{equation}
 T \simeq \frac{2}{\sqrt{\pi}}\, c\,(x^0)^{2+\alpha}x^1\,,\qquad
  \dot{T}\simeq\frac{2(2+\alpha)}{\sqrt{\pi}}\,c\,
  (x^0)^{1+\alpha}x^1\,,\qquad
  T'\simeq\frac{2}{\sqrt{\pi}}\, c\,(x^0)^{2+\alpha}\,.
\end{equation}
Of course, this satisfies leading contribution of
eq.(\ref{EoM1}),\footnote{The leading term of the equation of motion
is also satisfied at $x^1=0$.} and eq.(\ref{asympN0}) leads to
\begin{eqnarray}
 \VT \calF(z)&\simeq& 2c\,(x^0)^{2+\alpha}\,
  e^{-\frac{c^2}{\pi}(x^0)^{2(2+\alpha)}(x^1)^2}
  \to2\pi\,\delta(x^1)\,,\nn
 \VT\,\dot{T}^2\,\calF'(z)&\simeq&
  (2+\alpha)^2\,c\,(x^0)^\alpha(x^1)^2\,
  e^{-\frac{c^2}{\pi}(x^0)^{2(2+\alpha)}(x^1)^2}\to0\,.\label{zeropt}
\end{eqnarray}
{}From eqs.(\ref{region1}) and (\ref{zeropt}),
we find that the hamiltonian density asymptotically becomes
\begin{equation}
 \calH \to 2\pi T_{p}\,\delta(x^1)
    +\frac{T_p}{4\{f(x^1)\}^2}\,s(x^1)\,,\label{Hdensity}
\end{equation}
where
\begin{equation}
  s(y) = \left\{ \begin{array}{ll}
    1 & (y\ne0)\\ 0 & (y=0)\,.\end{array}\right.
\end{equation}
This shows the correct tension $2\pi T_p\, (=T_{p-1})$ of the
D$(p-1)$-brane at $x^1=0$. The non-zero energy density remains
and the pressure for
$x^0\to\infty$ is given by plugging eqs.(\ref{Tasym}) and
(\ref{region1}) into
\begin{equation}
 p=T^{11}= T_p\,\VT \left\{2(T')^2 \calF' - \calF\,\right\} \,.
\end{equation}
It is easy to check that the pressure falls off exponentially at late
time. Thus, this non-zero energy density turns out to be the tachyon
matter.

\section{Flow of energy}\label{sec:EF}
Having found the asymptotic solution, we consider the momentum
density, or the flow of energy in this section.
Eq.(\ref{EM}) leads to
\begin{equation}
 T^{01}=-2T_p~ \VT\, \dot{T}\, T'\, \calF'(z) \,.
\end{equation}
We see that $T^{01}=0$ at $x^1=0$. From eqs.(\ref{TP}), (\ref{eq:ep})
and (\ref{signf}), $\dot{T}$ and $T'$ at late time except for $x^1=0$
are given by
\begin{eqnarray}
 \dot{T} &\simeq& \sgn(x^1) \,, \\
 T' &\simeq& \sqrt{2\pi}\,\sgn(x^1) f'(x^1)\, \erfc(x^0/\sqrt{8})
	\,. \label{Tx1}
\end{eqnarray}
The remaining factor of $\vt\calF'$ at late time is given in the
previous section, so that it is easy to see that the energy ceases to
flow at $x^0\to\infty$, which is, of course, consistent with the
energy-momentum conservation.
Note that if $f'(x^1)=0$, or $f$ is constant, we cannot obtain
explicit form of $T'$ since eqs.(\ref{TP}) and
({\ref{eq:ep}}) is
correct up to $o\,(e^{-(x^0)^2/8})$. However, we see that
$T'$ is smaller than $\ep(x)$,
which does not alter the above result at $x^0\to\infty$.

Although the flow of the energy at early time is not given,
since we consider the time evolution of the kink, it is plausible that
$\dot{T}>0\  (x^1>0)$, $\dot{T}<0\  (x^1<0)$ and $T'>0$ in some
region around $x^1=0$ and hence the energy flows toward $x^1=0$ in
some region around there after some finite time, which means that the
energy does not dissipate to infinity.
We note that
the energy conservation implies $f(x^1)\sim 1/2$
if $f'(x^1)$ is small.

\section{Conclusion and discussion}
We have investigated the spatially inhomogeneous decay of an unstable
D-brane in boundary string field theory.
We have considered the equation of motion and
shown that $z$ must become either $-1$ or infinity as $x^0\to\infty$.
We have also given the time dependent kink solution of tachyon
eq.(\ref{TP}) with eqs.(\ref{eq:ep}), (\ref{signf}) at late time.
We found that the final state is the codimension one D-brane and
the tachyon matter, the pressure-less gas with non-zero energy density.
Furthermore, the energy (density) where the resultant codimension
D-brane exists is $2\pi T_p$ which exactly corresponds to the tension
of the codimension one D-brane, $T_{p-1}=2\pi T_p\,$.
Recently the inhomogeneous rolling tachyon of a different profile has
been analyzed in CFT approach \cite{LNT}. It is interesting to study
the relation to our analysis.

Finally we comment on the excitation of gauge fields.
In \cite{IU}, gauge fields in the rolling tachyon was studied.
We can perform a similar analysis with the rolling tachyon studied
here and we find that the excitations of the gauge field can
exist only on the lower-dimensional D-brane, while there are no
excitations on the remaining tachyon matter.

\vspace{5mm}
{\bf Acknowledgments:}
The work of SU is supported in part by the Grant-in-Aid for Scientific
Research No.13135212.

%%%%%%%%%%%%%%%%%%%%%%%%%%%%%

\end{document}